\documentclass{article}
\usepackage{spconf,amsmath,graphicx}

\usepackage{multirow}
\usepackage{numprint}
\usepackage{hyperref}
\usepackage{diagbox}
\usepackage{graphicx}
\usepackage{color}

\usepackage{algorithm}
\usepackage{algorithmic}
\newcommand{\rnum}[1]{\uppercase\expandafter{\romannumeral #1\relax}}

\title{Text adaptation for speaker verification with speaker-text factorized embeddings} 
\name{Yexin Yang$^\dagger$, Shuai Wang$^\dagger$, Xun Gong, Yanmin Qian, Kai Yu\thanks{$\dagger$: These authors have contributed equally to this work}
\thanks{Yanmin Qian and Kai Yu are the corresponding authors}
\thanks{Authors would like to thank Johan Rohdin for providing phoneme labels}
\thanks{This work has been supported by the China NSFC project No. U1736202. Experiments have been carried out on the PI supercomputer at Shanghai Jiao Tong University}}
\address{
        MoE Key Lab of Artificial Intelligence \\
    SpeechLab, Department of Computer Science and Engineering \\
    Shanghai Jiao Tong University, Shanghai, China \\ {\small \{yangyexin, feixiang121976, gongxun, yanminqian, kai.yu\}@sjtu.edu.cn}}

\begin{document}
\ninept
\maketitle
\begin{abstract}
Text mismatch between pre-collected data, either training data or enrollment data, and the actual test data can significantly hurt text-dependent speaker verification (SV) system performance. Although this problem can be solved by carefully collecting data with the target speech content, such data collection could be costly and inflexible. In this paper, we propose a novel text adaptation framework to address the text mismatch issue. Here, a speaker-text factorization network is proposed to factorize the input speech into speaker embeddings and text embeddings and then integrate them into a single representation in the later stage. Given a small amount of speaker-independent adaptation utterances, text embeddings of target speech content can be extracted and used to adapt the text-independent speaker embeddings to text-customized speaker embeddings. Experiments on RSR2015 show that text adaptation can significantly improve the performance of text mismatch conditions.

\end{abstract}

\begin{keywords}
speaker verification, text-dependent, text mismatch, adaptation
\end{keywords}

\section{Introduction}
\label{sec:intro}
Speaker verification aims to verify the client's claimed identity based on his/her speech. Considering the constraint on the speech content, speaker verification can be classified into two categories: text-dependent and text-independent. The former task requires the same speech content for the enrollment and test utterances, while the latter doesn't pose such a requirement, giving users more flexibility.

% Intuitively, for the text-independent speaker verification, we want to suppress the variability of speech content (phoneme information), meanwhile it's helpful to consider such information when building text-dependent systems.
For the text-independent speaker verification task, the speaker embedding extractor is usually trained on a large amount of unconstrained speech data, the text information is implicitly normalized, which is beneficial since the final speaker embeddings should get rid of the phonetic variability. Despite the good performance on the text-independent task \cite{snyderx,wang2019usage,huang2018angular,huang2018joint,zhang2017end}, directly applying the same model to the text-dependent task is problematic, for which text information is important. The common method to address such performance degradation is to collect training data which has the same speech content with the evaluation data, and this approach is usually adopted by companies for the wake-up word based speaker verification \cite{variani2014deep,heigold2016end,zhang2016end,zhang2019seq2seq,rahman2018attention}. However, recollecting application specific training data can be very expensive and inflexible.
% For example, the training speech segments in \cite{variani2014deep,heigold2016end} and \cite{zhang2016end} are collected from a large amount of speakers sharing the same content ``OK, Google'' or ``Hey, Cortana''. 
% The second approach is to incorporate the phoneme information into the modelling process. For instance, based on Google's d-vector framework\cite{variani2014deep}, we proposed j-vector, a multi-task learning framework in \cite{liu2015deep} to train the neural network to be both speaker-discriminative and phoneme-discriminative. Since collecting data for each specific application is very expensive, the second approach is more feasible and can take the advantage of a large amount of unconstrained speech data.

In real applications, the challenge not only comes from the text mismatch from the training data and evaluation data but also from the text mismatch between the enrollment and test data in the evaluation. For example, it's common in real applications where users would like to use multiple keywords to wake up smart devices. For example, Google devices allow ``OK Google'' and ``Hey Google'' \cite{rahman2018attention,wan2018generalized}. Some applications even involve more different keywords. 

In this paper, to avoid recollecting a large amount of application specific training data, we consider these two kinds of text mismatch in the ``text-adaptation framework'', in which text-independent speaker embeddings are adapted to customized text-dependent speaker embeddings according to a specific input.  We proposed a speaker-text factorization network which contains four parts: one generic feature learner, a speaker sub-net for text-independent ``spk'' embedding extraction, a text sub-net for ``text'' embedding extraction, and a ``combination'' sub-net to learn a text-adapted representation based on the information provided by ``text'' embedding. 

Different evaluation sets considering different types of text mismatch are derived from the RSR2015 \cite{larcher2014text} dataset, for the traditional text-dependent task where the text mismatch only exists between the training and evaluation data, the ``text ''embedding is computed from the same utterance as the ``spk'' embedding. For the case where text mismatch also appears between the enrollment and test data, we collected a very small amount of utterances from arbitrary speakers (different from the evaluation set) to compute the text embedding and adapt the enrollment speaker embeddings. Experimental results show a remarkable performance improvement for both conditions. Furthermore, the ``spk'' embedding extracted from the speaker sub-net with no text adaptation also outperforms the original x-vector baseline on the standard Voxceleb evaluation set.

\section{Related work}

\subsection{x-vector}
\label{sec:xvector}
 x-vector \cite{snyderx,snyder2017deep} is a time-delay neural network (TDNN) based speaker embedding learning framework. The model contains several frame-level time-delay layers, followed by a statistics pooling layer which aggregates the frame-level representation into a single segment-level representation. One or more embedding layers after the pooling layer can be incorporated in the segment-level layers to extract speaker embeddings. 
%  In our experiments, the x-vector extractor is used as the baseline and the backbone for our proposed framework.

\subsection{Segment-level phonetic label definition}
\label{sec:interspeech}
Researchers have investigated integrating the phonetic information into the speaker modeling process, most of which follow the frame-level multi-task learning paradigm \cite{liu2015deep,phonetic2018}.  In our previous work, we proposed a framework to consider phonetic information at the segment level, which is more compatible with the segment-level trained x-vector. The key point is how to define the phonetic labels for one speech segment, with multiple phonemes involved. We adopted a naive way as follows: for a given segment $\mathbf{x}$ with $N$ frames, the corresponding segment-level phoneme label $\mathbf{y}^t$ is represented as 
 
%  \begin{align*}
%      & \mathbf{y}^p = \{y_{1}, y_{2},\dots, y_{C}\} \\
%      & y_{c} = \frac{N_c}{N}
%  \end{align*}
\vspace{-17pt}
\begin{align*}
    \mathbf{y}^t = \{y_{1}, y_{2},\dots, y_{C}\}, \quad y_{c} = \frac{N_c}{N}
\end{align*}
where $C$ is the size of the chosen phoneme set. $N_c$ denotes the number of occurrences of the $c$-th phoneme in $\mathbf{x}$.

\section{Text-adaptation framework}
A typical deep speaker verification task involves three phases:
\begin{itemize}
    \item Training: the speaker embedding extractor is trained with a large amount of pre-collected data.
    \item Evaluation
        \begin{itemize}
            \item Enrollment: new speakers are enrolled by generating speaker embeddings via the well-trained extractor.
            \item Test: each test utterance is evaluated using the enrolled model of the claimed identity to make the verification decision.
        \end{itemize}
\end{itemize}

For the text-independent task, we don't pose any requirement on the text match either between the training and evaluation data or between the enrollment and test data. 

For the traditional text-dependent task, directly applying the systems trained for the text-independent speaker verification task usually achieves very poor performance due to the text mismatch between the training and evaluation data. The current state-of-the-art text-dependent speaker verification systems share the same methods with the text-independent ones, while the training data are collected for the customized application. For example, the training speech segments in \cite{variani2014deep,heigold2016end} and \cite{zhang2016end} are collected from a large amount of speakers sharing the same content ``OK Google'' or ``Hey Cortana''. Despite the good results achieved using this approach, it's expensive and inflexible to recollect the training data for each different phrases. 

Moreover, real-world applications usually don't follow the standard text-dependent regime. There are also scenarios where text mismatch also exists between enrollment and test utterances. For instance, it's common in some conditions users would like to use multiple keywords to wake up smart devices. For example, Google devices support both ``OK Google'' and ``Hey Google'' simultaneously \cite{rahman2018attention,wan2018generalized}. Some applications may require even more different keywords. Is it possible to allow the user only to enroll one of them and test on other different keywords?

% The motivation of this paper is to take advantage of the large amount of unconstrained data.

% In our previous work \cite{wang2019usage}, our main focus is on the text-independent speaker verification task, based on which we proposed the segment-level adversarial training to explicitly suppress the phoneme variability in the speaker embeddings. 
% In this paper we would like to extend the framework to factorize speaker and phonetic information, which allows us to better handle the text-dependent task and furthermore the text-adaptation task.
% In our previous work \cite{wang2019usage}, our main focus is on the text-independent speaker verification task, based on which we proposed the segment-level adversarial training to explicitly suppress the phoneme variability in the speaker embeddings. In this paper we would like to extend the framework to include the text-dependent task, we proposed the unified net which is depicted in Figure \ref{fig:system}.

\subsection{Text-adaptation for speaker embeddings}
\label{sec:text-adapt}
% mismatch affects performance, asr, precollected data (train data, enroll data)
To summarize the problems mentioned above, we would like to address the following two text-mismatch conditions:
\begin{itemize}
    \item text mismatch exists between the training and evaluation data, which is the traditional text-dependent task.
    \item text mismatch exists not only between the training and evaluation data but also between the enrollment and test data.
\end{itemize}

In the case that the pre-collected training data share the same text with the evaluation data, the text information is implicitly modeled in the speaker extractor. However, to address the two types of text mismatch mentioned above without the recollection of a huge amount of training data, the text information modeling should be explicitly considered. In this paper, we proposed a framework in which text-independent speaker embeddings could be adapted to text-dependent speaker embeddings, while the text information could be customized according to the input.

% Model performance is highly related to precollected data. Take speech recognition for example, the speaker identity mismatch between precollected training data and evaluation data may lead to huge performance degradation\cite{???}. The mismatch is even more complicated in text-dependent speaker verification task, where there are two types of mismatch: 1) between training data and evaluation data; 2) between enrollment data and evaluation data.

% To compensate the mismatch, traditional text-dependent speaker verification systems require same content as the test utterance (e.g. the desired wake-up word) in training phase and enrollment phase\cite{???}. However, 

\subsection{Speaker-text factorization network}

% \begin{table*}[t]
%     \caption{Validation Experiments on Voxceleb 1 Evaluation Set}
%     \label{tab:voxceleb_result}
%     \centering
%     \resizebox{0.7\textwidth}{!}{%
%     \begin{tabular}{cccccccc}
%         \hline
%         \multicolumn{2}{c}{System Configuration} & \multicolumn{2}{c}{Voxceleb1\_O} & \multicolumn{2}{c}{Voxceleb1\_E} & \multicolumn{2}{c}{Voxceleb1\_H} \\ \cline{3-4} \cline{5-6} \cline{7-8}
%         Architecture & Embedding Type & EER & minDCF  & EER & minDCF & EER & minDCF  \\ \hline
%         TDNN & spk & 2.888& 0.3281 & 3.055 & 0.3272 & 5.026 & 0.4646 \\ \hline
%         \multirow{2}{*}{Unified Net} & spk & \textbf{2.595} & \textbf{0.2940} & \textbf{2.784} & 0.2990 & \textbf{4.703} & 0.4292 \\
%          & spk + text & 2.808 & 0.2974 & 2.979 & \textbf{0.2988} & 4.905 & \textbf{0.4248}
%          \\ \hline
%     \end{tabular}%
%     }
% \end{table*}

\begin{figure}[!ht]
	\centering
	\includegraphics[width=0.8\linewidth]{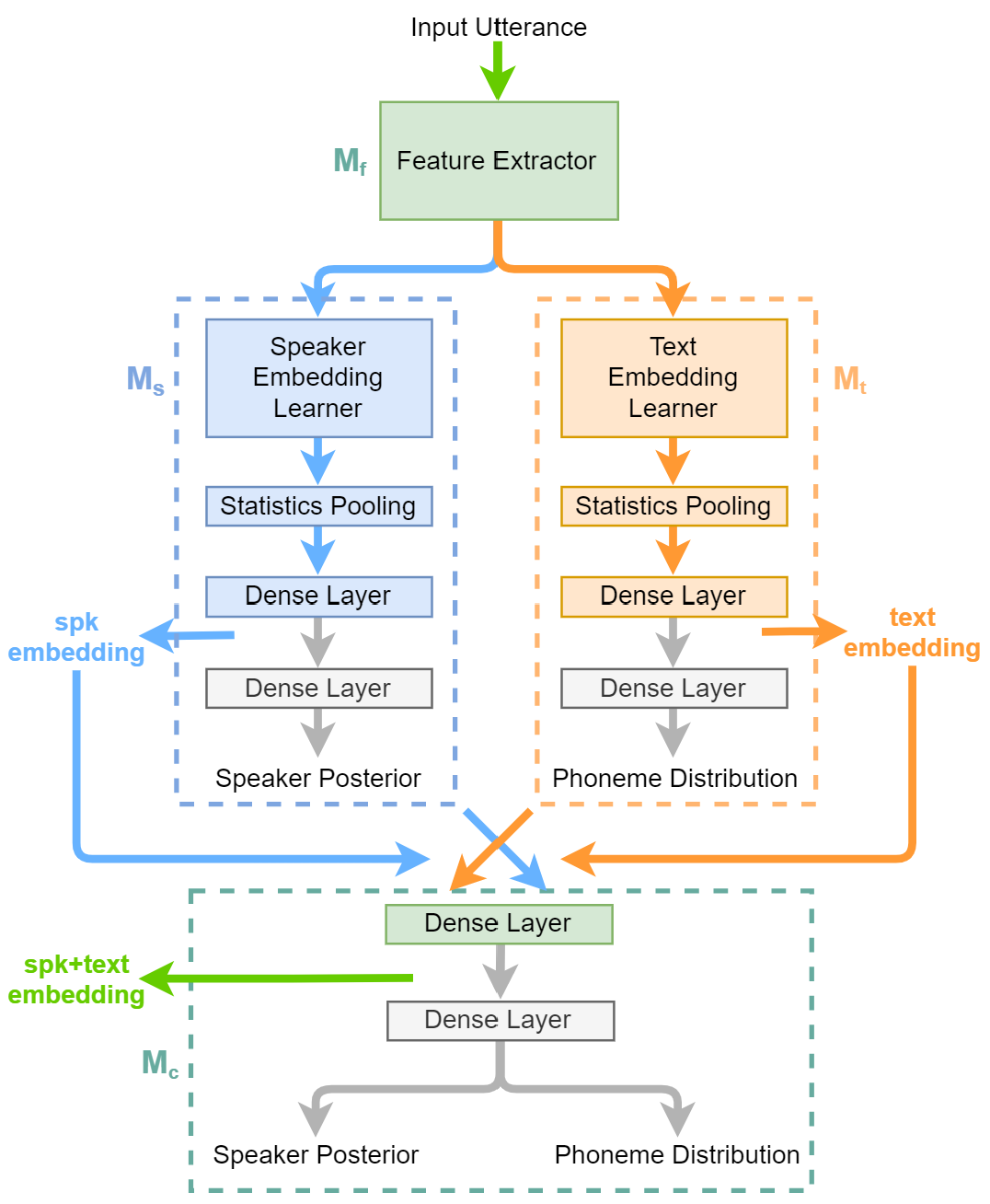}
	\caption{The proposed speaker-text factorization network.}
    \label{fig:system}
\end{figure}

As depicted in Figure \ref{fig:system}, The proposed model contains four parts: generic feature extractor $M_f$, two parallel sub-nets $M_s$ and $M_t$ for speaker discrimination and phoneme distribution learning respectively, and the ``combination'' sub-net $M_c$ for integrating both speaker and phonetic information. Following \cite{wang2019usage}, phoneme classifier $M_t$ predicts the normalized categorical occurrences of phonemes in one input segment and speaker classifier $M_s$ is a standard one predicting speaker classes. The speaker embedding $\text{ebd}_s$ and phoneme based text embedding $\text{ebd}_t$ extracted from $M_s$ and $M_t$ are then concatenated as the input into the combination network $M_c$, aiming to recover both the speaker identity and phonetic information. The model is trained jointly. Given the features of one training segment pair [$\mathbf{x}_s,\mathbf{x}_t$] and the corresponding speaker label $\mathbf{y}^s$ and phoneme label $\mathbf{y}^t$, the loss is defined as  $\mathcal{L}_{total}=\mathcal{L}_{s1}+\mathcal{L}_{t1}+\mathcal{L}_{s2}+\mathcal{L}_{t2}$, where
% Not a good way!!! M_{cs}, M_{cp} is not even mentioned!!!
\begin{align*}
    &\mathcal{L}_{s1} = \text{CE}(M_s(M_f(\mathbf{x}_s)), \mathbf{y}^s) \\
    &\mathcal{L}_{t1} = \text{KLD}(M_t(M_f(\mathbf{x}_t)), \mathbf{y}^t) \\
    &\mathcal{L}_{s2} = \text{CE}(M_c([\text{ebd}_s,\text{ebd}_t]), \mathbf{y}^s) \\
    &\mathcal{L}_{t2} = \text{KLD}(M_c([\text{ebd}_s,\text{ebd}_t]), \mathbf{y}^t)
\end{align*}
``spk'' embedding $\text{ebd}_s$ is computed from $\mathbf{x}_s$ using the speaker sub-net, while the ``text'' embedding $\text{ebd}_t$ is from $\mathbf{x}_t$ using the text sub-net. 
To better decouple speaker and phonetic information, the training pair [$\mathbf{x}_s,\mathbf{x}_t$] is randomly sampled from the training data, which can be identical or two utterances from two different speakers.

The ``spk'' embeddings extracted from the speaker sub-net are directly used for the text-independent task since no text adaptation is needed. For the traditional text-dependent speaker verification task, where the text embedding could be accurately computed for the enrollment and test data, the ``text'' embedding comes from the same utterance as the ``spk'' embedding. For the other scenario where text-mismatch exists from the enrollment and test utterance in the target trials, we use a very small amount of pre-collected data with the target text (e.g., 10 utterances from other speakers) to compute the ``text'' embedding and then use it to adapt the ``spk'' embeddings computed from the enrollment utterances, while the genuine ``text'' embeddings are used for the test utterances.

\section{Experimental setups}

\subsection{Data}
The Voxceleb and RSR2015 datasets are used in our experiments. All the phoneme labels are generated by a phoneme recognizer. More details could be referred to \cite{wang2019usage}.
Details about training and evaluation data preparation for the text-adaptation and text-independent evaluations will be given below, and the trial definitions are available online to reproduce our results on the customized RSR2015 evaluation set\footnote{ \url{https://github.com/Xflick/RSR2015_trials}}.

\vspace{-4pt}
\subsubsection{Training set}
For our experiments, Voxceleb2 development set is used for training the neural network and the Probabilistic Linear Discriminant Analysis (PLDA) back-end. This set contains 5994 speakers with 1092009 utterances. To train the neural network, we follow the data preparation process in Kaldi Voxceleb recipe, which cut the utterances to segments with length ranging from 2s to 4s. It should be noted that, unlike the recipe, we didn't use any data augmentation.

\subsubsection{Text-adaptation evaluation set}\label{sec:tarsr}
Two individual evaluation sets are created corresponding to different text-mismatch cases:

\noindent\textbf{Mismatch between training and evaluation data}: When the mismatch is only between training and evaluation data, it becomes the traditional text-dependent task. The evaluation set is derived from the evaluation portion of RSR2015 \cite{larcher2014text} Part I. It contains 30 fixed phrase utterances of 3-4s duration from 106 speakers (57 males, 49 females). 
Every phrase is spoken 9 times by each speaker, 3 of which are taken for registering, and the rest are used for testing. As shown in Table \ref{tab:trial}, for a standard text-dependent task, there are four possible types of trials, among which TC represents the target trials and TW, IC, IW denote three non-target conditions.

% \vspace{-\baselineskip}
\begin{table}[!htb]
\caption{ Types of trials for the text-dependent task}
\centering
\begin{tabular}{ c c c }
\hline
  & Correct Content   &   Wrong Content   \\ \hline
Target & \textit{TAR-correct} (TC) & \textit{TAR-wrong} (TW)\\ 
Impostor &\textit{IMP-correct} (IC) &    \textit{IMP-wrong} (IW)   \\ \hline
\end{tabular}
\label{tab:trial}
\end{table}

Since in the original trial definition, about $90\%$ of the original trials belong to the very easy IW case, we generate our own trial list following the ratio: TC:TW:IC:IW=1:3:3:3.

\noindent\textbf{Mismatch between enrollment and test data}: As mentioned before, there are 30 different fixed phrases in the RSR2015 part1 evaluation set. We randomly select ten of them and generate ten evaluation subsets. Two enrollment conditions are considered, the text-independent and text-dependent. For the former condition, the text of three enrollment utterances is randomly selected, while for the latter one, the text is shared among the enrollment utterances and has no overlap between the text of test utterances.
% In both conditions the text of enrollment and test utterances will have no overlap.
The text embedding for adaptation is computed using 10 random utterances from the development set with the same text as target text (speakers are other from the evaluation set). 
To better exhibit the text awareness of our speaker-text factorization network, we increase the number of TC trials. And the trial list for this specific task follows the ratio: TC:TW:IC:IW=1:1:1:1.

\subsubsection{Text-independent evaluation set}
\textbf{Voxceleb1 evaluation set}: To validate the performance of our baseline and proposed system, we first report the results on the official Voxceleb1 evaluation set. The cleaned trial lists are used:  VoxCeleb1 (denoted as Voxceleb1-O, O for ``original''), Voxceleb1-E (extended), and cleaned Voxceleb1-H (hard).

\noindent\textbf{RSR2015 evaluation set}: 
To better exhibit the effectiveness of our proposed framework, we designed one more text-independent evaluation set based on RSR2015, all the trial pairs are the same as the ones defined for traditional text-dependent case in Section \ref{sec:tarsr}, while the trials from TW condition are now treated as target.
% \begin{table}
% \centering
% \begin{tabular}{|c|c|c|}
% \hline
% \diagbox{Utterance}{Speaker} & match & not match \\
% \hline
% match & Target & Nontarget \\
% \hline
% not match & Nontarget & Nontarget \\
% \hline
% \end{tabular}
% \caption{trials}
% \label{tab:trials}
% \end{table}

% As a result, minDCF0:01 cannot be reliably estimated
% with the relatively few non-target trials available in the test set
% (please see Figure 5 and its caption). We therefore report results
% also on minDCF0:1.

% \subsection{The phoneme recognizer}
% \label{sec:phoneme}
% The frame-level phoneme labels are generated using the official Kaldi \cite{povey2011kaldi} Tedlium speech recognition recipe ({\texttt{s5\_r3}}).  This recipe uses a TDNN based acoustic model with i-vector adaptation and a RNN based language model. Phoneme posteriors are obtained from the lattices via the forward-backward algorithm and then converted to hard labels.  There are 39 phonemes, each coming in  four different versions depending on their position in the word, plus a silence (SIL) and noise class (NSN) that has 5 versions each, resulting in 166 \emph{phoneme classes}.

\subsection{System configurations}
Standard x-vector \cite{snyderx} system with five time delay layers and two dense layers is used as our baseline system.
The proposed Factorization Net system is modified from the baseline system, with three time delay layers extracting generic features. $M_s$ and the $M_t$ have identical structures, both having two time delay layers, one statistics pooling layer and two dense layers, except that one is for speaker classification and the other is for phonetic information prediction. $M_c$ has two dense layers in common and two output layers for two tasks. It is notable that when extracting only \textit{speaker embedding}, the Factorization Net has exactly the same structure as the baseline TDNN model.

40-dimensional Fbank features are used for model training. 
The neural networks are trained on 4 GPUs with a batch size of 256. 
Stochastic gradient descent with learning rate 0.01, momentum 0.9 and weight decay 1e-4 is used to optimize the model.
Batch normalization is applied after ReLU activation function.
% ReLU function is used to activate all neural networks after batchnorm.

PLDA is applied on Voxceleb1 evaluation set to validate the correctness of our system, for other evaluation sets, to get rid of the impact of PLDA compensation and focus on the properties of learned embeddings, and simple cosine scoring is utilized.
All architectures are implemented in PyTorch \cite{pytorch}. We report the performance of our models in terms of Equal error rate (EER) and Minimum detection cost (minDCF) with $P_{tar}$ set as 0.01.

% \subsubsection{Model setup}

% \textbf{Network \rnum{1}} has $M_f$ with three time delay layers to extract generic features, $M_s$ with two time delay layers, a statistics pooling layer and two dense layers and $M_t$ has the identical structure as $M_s$. $ebd_s$ is extracted from the first dense layer of $M_s$, $ebd_t$ is extract from the first dense layer of $M_t$.

% \textbf{Network \rnum{2}} has 2 dense layers in common. 
% After these, one dense layer is optional for both speaker and phoneme classifier. %TODO: optional or necessity?
% $ebd$ is extracted from common dense layers(both layer is considered).

%TODO: layer detail is needed ? or not? like LSTM ...

% \subsubsection{Backend}

% We use cosine score as backend for ...
% And PLDA model is used in evaluating ...

\section{Results and Analysis}
% NOTE: voxceleb1 is PLDA score
\begin{table*}[t]
    \caption{Validation experiments on Voxceleb 1 evaluation set}
    \label{tab:voxceleb_result}
    \centering
    \resizebox{0.7\textwidth}{!}{%
    \begin{tabular}{cccccccc}
        \hline
        \multicolumn{2}{c}{System Configuration} & \multicolumn{2}{c}{Voxceleb1\_O} & \multicolumn{2}{c}{Voxceleb1\_E} & \multicolumn{2}{c}{Voxceleb1\_H} \\ \cline{3-4} \cline{5-6} \cline{7-8}
        Architecture & Embedding Type & EER & minDCF  & EER & minDCF & EER & minDCF  \\ \hline
        TDNN & spk & 2.888& 0.3281 & 3.055 & 0.3272 & 5.026 & 0.4646 \\ \hline
        Factorization Net & spk & \textbf{2.595} & \textbf{0.2940} & \textbf{2.784} & \textbf{0.2990} & \textbf{4.703} & \textbf{0.4292} \\
        \hline
    \end{tabular}%
    }
\end{table*}
\subsection{Validation results on the text-independent task}
x-vector was proposed for the text-independent task, to show the correctness of the baseline model and the proposed model, and we first report the results on the standard Voxceleb1 evaluation set in Table \ref{tab:voxceleb_result}. Since we only used the clean training data, the baseline is quite strong compared to the ones in the literature \cite{chung2018voxceleb2,xiang2019margin,xie2019utterance}. As shown in Table \ref{tab:voxceleb_result}, the embedding extracted from the speaker sub-net of the proposed model reduces the EERs on Voxceleb\_O, Voxceleb\_E and Voxceleb\_H from 2.888\%, 3.055\% and 5.026\% to 2.595\%, 2.784\% and 4.703\%, respectively. A similar improvement can also be observed in terms of minDCF. 
For the experiments on the RSR2015 text-independent evaluation set, similar performance improvement for the speaker embedding from the speaker sub-net is observed.
% However, we found that adding text information to the embedding leads to huge performance degradation. Compared to the Voxceleb test utterances, the ones in RSR2015 are much shorter, which means the phonetic variability is more harmful when we consider it as a text-independent task.

% As mentioned in Section \ref{sec:sampling}, the proposed training data sampling algorithm requires the neural network to recover the speaker identity and text information from different utterances in the recombination stage, which achieves similar impact on the speaker and text sub-nets as the adversarial training in our previous work\cite{}.
% \vspace{-\baselineskip}
% \begin{table}[!htb]
% \centering
% \caption{Experiments on RSR2015 text-independent evaluation set}
% \label{tab:rsr_tid_result}
% \begin{tabular}{cccc}
% \hline
% \multicolumn{2}{c}{System Configuration} & \multirow{2}{*}{EER (\%)} & \multirow{2}{*}{minDCF} \\
% Architecture & Embedding Type &  &  \\ \hline
% TDNN & spk & 7.220 & 0.7068 \\ \hline
% \multirow{2}{*}{Factorization Net} & spk & \textbf{6.239} & \textbf{0.6721} \\
%  & spk+text & 14.92 & 0.7750 \\ \hline
% \end{tabular}
% \end{table}

% \vspace{-\baselineskip}
\begin{table}[!htb]
\centering
\caption{Experiments on RSR2015 text-independent evaluation set}
\label{tab:rsr_tid_result}
\begin{tabular}{cccc}
\hline
\multicolumn{2}{c}{System Configuration} & \multirow{2}{*}{EER (\%)} & \multirow{2}{*}{minDCF} \\
Architecture & Embedding Type &  &  \\ \hline
TDNN & spk & 7.220 & 0.7068 \\ \hline
Factorization Net & spk & \textbf{6.239} & \textbf{0.6721} \\
\hline
\end{tabular}
\end{table}

% \vspace{-12pt}
\subsection{Text adaptation to address the text-mismatch problems}
\subsubsection{Mismatch between training and evaluation data}
As shown in Table \ref{tab:rsr_td_result}, when the mismatch happens between training and evaluation data, integrating text information into embedding significantly improves the performance. The EER of the system is reduced from $6.671\%$ to $1.542\%$, while the minDCF is reduced from $0.5234$ to $0.1246$.

% \vspace{-\baselineskip}
\begin{table}[!htb]
\centering
\caption{Experiments on RSR2015 text-adaptation evaluation set (mismatch between training and evaluation data)}
\label{tab:rsr_td_result}
\begin{tabular}{cccc}
\hline
\multicolumn{2}{c}{System Configuration} & \multirow{2}{*}{EER (\%) } & \multirow{2}{*}{minDCF} \\
Architecture & Embedding Type &  &  \\ \hline
TDNN & spk & 6.671 & 0.5234 \\ \hline
\multirow{2}{*}{Factorization Net} & spk & 6.010 & 0.5144 \\
 & spk+text & \textbf{1.542} & \textbf{0.1246} \\ \hline
\end{tabular}
\end{table}

Table \ref{tab:rsr_td_breakdown} shows the results when different error types are individually analyzed. The errors TW and IW, which are resulted from the wrong text, are greatly reduced as expected. The speaker error (IC) is also decreased from $1.919\%$ to $1.101\%$.

% \vspace{-\baselineskip}
\begin{table}[!htb]
\centering
\caption{EERs with regard to different error types on RSR2015 text-adaptation evaluation set (mismatch between training and evaluation data)}
\label{tab:rsr_td_breakdown}
\begin{tabular}{ccccc}
\hline
\multicolumn{2}{c}{System Configuration} & \multicolumn{3}{c}{EER (\%)} \\ \cline{3-5}
Architecture & Embedding Type & TW & IC & IW \\ \hline
TDNN & spk & 10.60 & 1.919 & 1.007 \\ \hline
\multirow{2}{*}{Factorization Net} & spk & 10.32 & 1.385 & 0.7867 \\
 & spk+text & \textbf{2.454} & \textbf{1.101} & \textbf{0.1573} \\ \hline
\end{tabular}
\end{table}

\subsubsection{Mismatch between enrollment and test data}
% As mentioned in Section \ref{sec:text-adapt}, text-adaptation task can be regarded as a generalized case of text-dependent task, where we loosen the constraint to allow arbitrary enrollment phrase.
As shown in Table \ref{tab:ta-result}, when the mismatch happens between not only training and test data, but also enrollment and test data, most systems fail on the task. However, by using the target text embedding instead of the genuine text embedding to adapt enrollment ``spk'' embedding, the system performance is substantially improved, which shows the effectiveness of our factorization net to generate text-customized speaker embeddings.

\begin{table}[!htb]
\centering
\caption{Text-independent/Text-dependent enrollment (introduced in Sec \ref{sec:tarsr}) EERs(\%) on RSR2015 text-adaptation evaluation set (mismatch between enrollment and test data)}
\label{tab:ta-result}
\resizebox{\linewidth}{!}{%
\begin{tabular}{c||c|ccc}
\hline
\multirow{2}{*}{Subset} & TDNN & \multicolumn{3}{c}{Factorization Net} \\ \cline{2-5} 
 & spk & spk & \begin{tabular}[c]{@{}c@{}}spk\\ +text\end{tabular} & \begin{tabular}[c]{@{}c@{}}spk\\ +adapt\_text\end{tabular} \\ \hline
1 & 27.34/29.21 & 27.05/28.31 & 26.42/26.29 & \textbf{12.26/15.07} \\
2 & 24.61/22.73 & 26.03/24.09 & 24.87/23.18 & \textbf{13.71/13.99} \\
3 & 22.18/24.33 & 23.41/25.37 & 20.75/21.49 & \textbf{10.84/14.49} \\
4 & 22.56/20.19 & 22.88/19.97 & 24.87/17.14 & \textbf{6.912/7.227} \\
5 & 28.16/32.45 & 26.74/30.99 & 27.93/34.84 & \textbf{10.51/16.98} \\
6 & 22.46/21.32 & 22.76/21.74 & 21.38/23.40 & \textbf{9.985/10.81} \\
7 & 22.20/25.09 & 22.02/25.23 & 22.02/27.67 & \textbf{7.898/12.52} \\
8 & 29.97/29.09 & 29.64/28.88 & 30.24/30.44 & \textbf{14.21/15.32} \\
9 & 22.86/24.73 & 22.43/25.04 & 22.91/24.66 & \textbf{9.450/14.56} \\
10 & 22.91/25.50 & 22.51/25.34 & 23.15/25.76 & \textbf{9.083/12.35} \\ \hline
avg & 24.53/25.46 & 24.55/25.50 & 24.45/25.49 & \textbf{10.49/13.33} \\ \hline
\end{tabular}%
}
\end{table}

% \begin{table}[]
% \begin{tabular}{ccccc}
% \hline
% \multicolumn{2}{c}{System Configuration} & \multicolumn{3}{c}{EER (\%)} \\ \cline{3-5} 
% Architecture & Embedding Type & TW & IC & IW \\ \hline
% TDNN & spk & 48.93/50.91 & 4.395/6.013 & 4.528/6.652 \\ \hline
% \multirow{3}{*}{Unified Net} & spk & 48.96/51.38 & 3.886/5.213 & 3.927/5.703 \\
%  & spk+text & 48.38/51.68 & 5.336/6.664 & 5.530/8.731 \\
%  & spk+adapt\_text & 17.76/23.02 & 4.646/6.033 & 1.544/2.197 \\ \hline
% \end{tabular}
% \end{table}

% \begin{table*}[]
% \centering
% \caption{Adapt}
% \resizebox{\textwidth}{!}{%
% \begin{tabular}{ccccccccccccc}
% \hline
% \multicolumn{2}{c}{System Configuration} & \multicolumn{10}{c}{Subset} & \multirow{2}{*}{Average} \\ \cline{3-12}
% Architecture & Embedding Type & 1 & 2 & 3 & 4 & 5 & 6 & 7 & 8 & 9 & 10 &  \\ \hline
% TDNN & spk & 27.34 & 24.61 & 22.18 & 22.56 & 28.16 & 22.46 & 22.2 & 29.97 & 22.86 & 22.91 & 24.53 \\ \hline
% \multirow{3}{*}{Unified Net} & spk & 27.05 & 26.03 & 23.41 & 22.88 & 26.74 & 22.76 & 22.02 & 29.64 & 22.43 & 22.51 & 24.55 \\
%  & spk+text & 26.42 & 24.87 & 20.75 & 24.87 & 27.93 & 21.38 & 22.02 & 30.24 & 22.91 & 23.15 & 24.45 \\
%  & spk+adapt\_text & 12.26 & 13.71 & 10.84 & 6.912 & 10.51 & 9.985 & 7.898 & 14.21 & 9.450 & 9.083 & 10.49 \\ \hline
% \end{tabular}%
% }
% \end{table*}

% \vspace{-5pt}
\section{Conclusions and future works}
Text mismatch between the training data and evaluation data can lead to huge performance degradation for the text-dependent speaker verification. One common solution is to collect application-specific training data that share the same text information as the evaluation data. To get rid of the expensive and inflexible data collection process and take advantage of the large amount of unconstrained speech data, we proposed a ``text-adaptation speaker verification'' framework, in which the text-independent speaker embeddings could be adapted to text-customized ones according to the specific adaptation input. A speaker-text factorization network is proposed, which first factorizes a speech segment into a text-independent speaker embedding and a speaker-independent text embedding and then recombines them as one single embedding containing both information. We first verify the proposed method without text adaptation on standard text-independent Voxceleb evaluation set and observe consistent performance improvement on all the three trial lists. Results on three customized evaluation sets derived from the RSR2015 dataset show that the proposed method using text adaptation can greatly reduce the errors caused by the text-mismatch between the training and evaluation data and between the enrollment and test data. 

In the future work, we will make more efforts to allow the model to utilize simple plain text instead of the text embedding computed from specific audios for the text adaptation on speaker embeddings.

% \newpage
\bibliographystyle{IEEEbib}
\bibliography{strings,refs}

\end{document}